\documentclass[a4paper,11pt]{article}
\usepackage{pos}

\usepackage{multirow}
\newcommand{\abs}[1]{\lvert#1\rvert}

\title{Proton charge radius from a dispersive analysis of the experimental data over the space- and time-like regions}
\ShortTitle{Proton charge radius from the dispersion theoretical analysis}

\author*[a]{Yong-Hui Lin}
\affiliation[a]{Helmholtz-Institut~f\"{u}r~Strahlen-~und~Kernphysik~and~Bethe~Center~for	Theoretical~Physics, Universit\"{a}t~Bonn, \\ D-53115~Bonn,~Germany}
\emailAdd{yonghui@hiskp.uni-bonn.de}

\abstract{We present a dispersion theoretical analysis of the experimental data on the electromagnetic form factors of the nucleon covering both the space- and time-like regions. The nucleon form factors over the full range of momentum transfers and the nucleon radius are extracted with high precision. The statistical uncertainties of the extracted form factors and radius are estimated using the bootstrap method, while systematic errors are determined from variations of the spectral functions. For the proton charge radius, we find $r_E^p = 0.840^{+0.003}_{-0.002}{}^{+0.002}_{-0.002}$~fm, in line with previous analyses of spacelike data alone and also the muonic Lamb shift determination.}

\FullConference{%
  The 10th International Workshop on Chiral Dynamics - CD2021\\
  15-19 November 2021\\
  Online
  }

\begin{document}
\maketitle

\section{Introduction}
The most fundamental approach to explore the intrinsic structure of a particle in experiments is to scatter on it. During the last decade, the electromagnetic structure of the proton has been becoming a hot topic in hadron and nuclear physics, which arouses great interest and efforts both in experimental and theoretical research. Such a research upsurge is triggered by the well-known ``proton radius puzzle'', which appears since the first measurement on the proton electric radius from the muonic hydrogen, which led to the so-called small radius, $r_p= 0.84184(67)\,$fm, with unprecedented precision but differing by 5$\sigma$ from the CODATA value, $0.8775(51)\,$fm, at that time, was reported by Ref.~\cite{Pohl:2010zza} in 2010. See Refs.~\cite{Carlson:2015jba,Hammer:2019uab,Karr:2020wgh} for recent reviews. The proton charge radius, defined as $r_p^2= -6 dG_{E,p}(Q^2)/dQ^2|_{Q^2=0}$, can be evaluated straightforwardly from the Sachs form factors of the proton which contain all information about the electromagnetic structure of proton and are accessed by elastic lepton-proton scattering. Recently, a new set of $e$-$p$ scattering data that reaches extremely low momentum transfer, around $10^{-4}$~GeV$^2$, was reported by the PRad collaboration~\cite{Xiong:2019umf}. Together with the previous MAMI-C's measurement~\cite{A1:2013fsc}, it serves as a good platform where one can extract the proton form factors in the space-like region, specifically $0.000215<Q^2<0.977$~GeV$^2$ with $t=-Q^2<0$ for the space-like transfer momentum squared. On the other hand, the time-like form factors of the proton ($t>0$) are embedded into the cross sections of the annihilation process $e^+e^-\to p\bar{p}$. Starting from the earliest measurement performed by the ADONE73 group in 1973~\cite{Castellano:1973wh} to the very recent experiment by BESIII in 2021~\cite{BESIII:2021rqk}, there is a huge amount of annihilation data that covers the energy region of $3.52<t<20.25$~GeV$^2$ has been accumulated. Especially, those data from BABAR~\cite{BaBar:2013ves} and BESIII~\cite{BESIII:2021rqk} also include the modulus of the form factor ratio of the proton, $|G_E^p/G_M^p|$, and the differential cross sections at some energy points. Motivated by the rich scattering and annihilation database, we would like to implement a combined analysis to extract the nucleon form factors over the space-like and time-like region.

Dispersion theory is the best candidate of theoretical tools when one explores the physics over the large energy region due to the following advantages. First of all, it is based on the unitarity and analyticity which are two fundamental properties for a physical observable required in the quantum field theory. Secondly, one can relate the time-like physics to the space-like ones by means of the crossing symmetry in the dispersion theory. And finally, when specific to the nucleon form factors, it is convenient to include the strictures from perturbative QCD at very large momentum transfer (see the recent review~\cite{Lin:2021umz} for more details). Here, we extend the previous applications of the dispersion theory in the extraction of nucleon form factors from the space-like scattering data~\cite{Lin:2021umk,Lin:2021umz} to include the time-like annihilation data. It leads us to the best phenomenological nucleon form factors which incorporate all the physical knowledge we have so far in the theoretical aspect and are consistent with all existing scattering and annihilation data in the experimental aspect. With such powerful nucleon form factors, we also obtain a high-precision determination of the proton electric and magnetic radii and the neutron magnetic radius.

\section{Formalism}
At the first order in the electromagnetic fine-structure constant $\alpha$, namely the Born-approximation, the differential cross section for the elastic electron-proton scattering, $e\,(p_1) + p\,(p_2) \to e\,(p_3) + p\,(p_4)$, can be expressed through the Sachs form factors (FFs) as (working in the laboratory frame, namely the rest target frame)
\begin{equation}\label{eq:xs_ros}
	\frac{d\sigma}{d\Omega} = \left( \frac{d\sigma}
	{d\Omega}\right)_{\rm Mott} \frac{\tau}{\epsilon (1+\tau)}
	\left[G_{M}^{2}(t) + \frac{\epsilon}{\tau} G_{E}^{2}(t)\right]\, ,
\end{equation}
where $\tau = -t/4m_p^2$ with $t$ the four-momentum transfer squared defined as $t=(p_1-p_3)^2$ and $m_p$ the proton mass, $\epsilon = [1+2(1+\tau)\tan^{2} (\theta/2)]^{-1}$ is the virtual photon polarization. $\theta$ is the scattering angle of outgoing electron in the laboratory frame, and $({d\sigma}/{d\Omega})_{\rm Mott}$ is the Mott cross section which corresponds to scattering off a point-like proton. Note that the nucleon FFs in elastic $ep$ scattering are often displayed as a function of $Q^2$ in literature since $t= -Q^2 < 0$ is spacelike. Considering the annihilation process $e^+\,(p_1)+e^-\,(p_2)\to p\,(p_3)+\bar{p}\,(p_4)$, one commonly takes the total cross section (working in the center-of-mass (c.m.) frame),
\begin{align}\label{eq:xs_time}
	\sigma_{e^+e^- \rightarrow p\bar{p}}(t) &= \frac{4\pi\alpha^2\beta}{3 t}C(t)
	\left[|G_M(t)|^2+\frac{2m_p^2}{t}|G_E(t)|^2\right]\notag\\
	&\equiv \frac{4\pi\alpha^2\beta}{3 t}C(q^2)\left(1+\frac{2m_p^2}{t}\right)|G_{\rm eff}^p(t)|^2.
\end{align}
This defines the effective form factor
$G_{\rm eff}$
\begin{align}
	\left|G_{\rm eff}\right| \equiv \sqrt{\frac{|G_E|^2+\xi|G_M|^2}{1+\xi}}~,
	\label{eq:geff}
\end{align}
where $\xi=t/(2m_p^2)$ with $t=(p_1+p_2)^2$ denotes the center-of-mass energy, $\beta=\sqrt{1- 4m_p^2/t}$ is the velocity of the proton and $\theta$ is the emission angle of the proton in the $e^+e^-$ c.m. frame. $C(t)=\frac{y}{1-e^{-y}}$ with $y=\frac{\pi\alpha}{\beta}$ is the Sommerfeld factor that accounts for the electromagnetic interaction between the outgoing proton and antiproton. Here, $t>0$ is time-like for the annihilation process. Eqs.~\eqref{eq:xs_ros} and \eqref{eq:xs_time} are the basic formulae to analyze the measured data from the $ep$ scattering and $e^+e^-$ annihilation, respectively. In practice, the time-like data is commonly given as the effective form factor Eq.~\eqref{eq:geff} directly. To construct the parametrization of nucleon FFs in the dispersion theoretical framework, it is convenient to rewrite the Sachs electric and magnetic FFs to the Dirac and Pauli FFs with the following combinations, 
\begin{equation}
	G_E(t) = F_1(t)-\tau F_2(t)~, ~~~~G_M(t) = F_1(t)+F_2(t)~,
\end{equation}
where the normalization of Dirac and Pauli FFs is given by,
\begin{equation}
	\label{eq:norm}
	F_1^p(0) = 1\,,  \; F_1^n(0) = 0\,,\; F_2^p(0) = \kappa_p \,,  \; F_2^n(0) = \kappa_n \, ,
\end{equation}
with $\kappa_p=1.793$ and $\kappa_n=-1.913$ in units of the nuclear magneton, $\mu_N = e/(2m_p)$.
And then decompose them into the pure isospin form
\begin{equation}
	F_i^s = \frac{1}{2} (F_i^p + F_i^n) \, , \quad
	F_i^v = \frac{1}{2} (F_i^p - F_i^n) \, ,
\end{equation}
where $i = 1,2 \,$. 

\begin{figure}[htbp]
	\centering
	\includegraphics[width=0.7\textwidth]{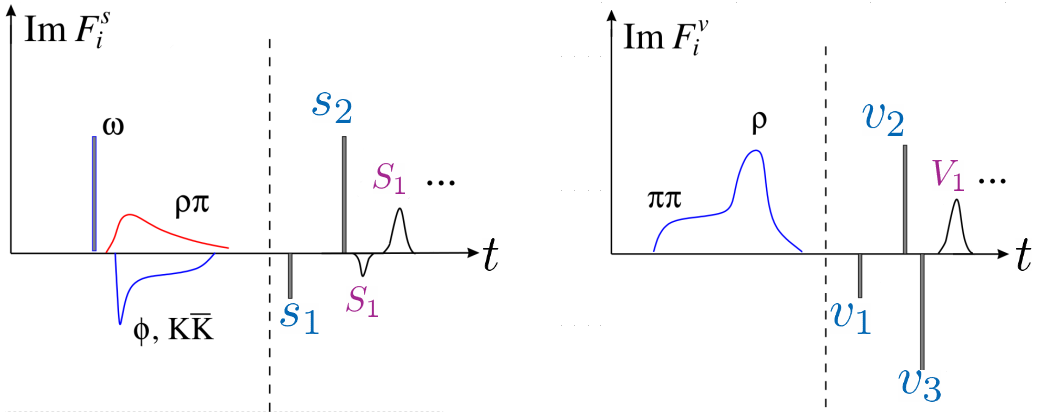}
	\caption{Cartoon of the isoscalar (left) and isovector (right) spectral function in terms of continua and (effective) vector meson poles. The vertical dashed line, located at $1~\mathrm{GeV}$, separates the well-constrained low-mass region from the high-mass region which is parameterized by effective poles.}
	\label{fig:cartoon}
	\vspace{-3mm}
\end{figure}

The dispersion relations for these isoscalar and isovector FFs read as\begin{equation}
	F_i(t) = \frac{1}{\pi} \, \int_{t_0}^\infty \frac{{\rm Im}\, 
		F_i(t')}{t'-t-i\epsilon}\, dt'\, \hspace{1cm}i = 1,2~,
	\label{eq:disprel}
\end{equation}
where $t_0 = 4M_\pi^2 \, (9M_\pi^2)$ denotes the threshold of the lowest cut for the isovector (isoscalar) case, with $M_\pi$ the charged pion mass. Eq.~\eqref{eq:disprel} states that the full nucleon form factor can be constructed from its imaginary part which incorporates all information concerning the analytical structure of the nucleon form factor. In principle, the imaginary part entering the integrals on the right side of Eq.~\eqref{eq:disprel} can be obtained completely from a spectral decomposition that is illustrated as~\cite{Chew:1958zjr,Federbush:1958zz}
\begin{equation}
		{\rm Im}\, \langle N(p') \overline{N}(\bar{p}') | j_\mu^{\rm em}(0) | 0 \rangle
		\propto \sum_\lambda
		\langle N(p') \overline{N}(\bar{p}') | \lambda \rangle
	\times
		\langle \lambda | j_\mu^{\rm em} (0) | 0 \rangle
		\,\delta^4(p'+\bar{p}'-p_\lambda)\,.
		\label{spectro}
\end{equation}
The first term on the right side indicates the matrix element for scattering of the intermediate states $|\lambda\rangle$ into a $N\bar{N}$ pair. And the second term is the matrix element for creation of the intermediate states. Both of these two terms can be estimated from the corresponding experimental measurements. In practice, the spectral decomposition, however, can be applied to estimate only the low energy part of the nucleon FFs due to the lack of relevant measurements in the high energy region. Thus, the vector meson dominance (VMD) model is introduced to get the high energy part of the nucleon FFs in our dispersion theoretical parametrization, see Ref.~\cite{Lin:2021umz} for more details. Here, we summarize the ingredients in the spectral functions,
\begin{itemize}
	\item In isoscalar sector: $\omega (782)$ and $\phi (1020)$ mesons,  $\pi\rho$ and $\bar{K}K$ continua, some effective narrow vector meson poles $s_1$, $s_2\cdots$, broad vector meson poles $S_1$, $S_2\cdots$.
	\item In isoscalar sector: the $\pi\pi$ continuum, some effective narrow vector meson poles $v_1$, $v_2\cdots$, broad vector meson poles $V_1$, $V_2\cdots$.
\end{itemize}
The specific expressions of our spectral functions can be found in Ref.~\cite{Lin:2021xrc}. A cartoon of the resulting (isoscalar and isovector) spectral functions is shown in Fig.~\ref{fig:cartoon}.

Now, we can fit to experimental data with the dispersion theoretical nucleon FFs. The fit parameters in our dispersive framework are the various vector meson residua $a_i^V$ and the masses of the effective vector mesons $s_i, v_i$. Note that the masses of $\omega (782)$ and $\phi (1020)$ are fixed at their physical values and all two-body continua is fixed as input. We enforce various constraints that refined from the physical knowledge we have about the nucleon FFs so far to reduce the number of parameters. They are listed as
\begin{itemize}
	\item The four normalization as given in Eq.~\eqref{eq:norm},
	\item Neutron charge radius squared fixed at recent determination from Ref.~\cite{Filin:2019eoe},
	\begin{equation}\label{eq:r2En}
		\langle r_n^2\rangle = -0.105^{+0.005}_{-0.006}~{\rm fm}^2~.
	\end{equation}
	\item Superconvergence relations which are consistent with the requirements of perturbative QCD at very large momentum transfer~\cite{Lepage:1980fj}.
	\begin{equation}
		\label{eq:scr}
		\int_{t_0}^\infty {\rm Im}\, F_i (t) \;t^n dt =0\, , \quad i = 1,2 \, ,
	\end{equation}
	\item Physical limit for the parameters: $0.5\,\mbox{GeV}^2<a_{1}^\omega < 1\,\mbox{GeV}^2$, $|a_{2}^\omega| < 0.5\,\mbox{GeV}^2$ and $|a_{1}^\phi| < 2\,\mbox{GeV}^2$, $|a_{2}^\phi| < 1\,\mbox{GeV}^2$, all other residua are bounded as $|a_i^V| < 5\,$GeV$^2$, the masses of the narrow effective poles are constrained in the range of $1~\mathrm{GeV}$-$2m_p$ and the masses of the broad effective poles are required to locate in the range of $2m_p$-$3~\mathrm{GeV}$.
\end{itemize}
The data sets included in our fits are listed in Table~\ref{tab:dbase}. Explicit references can be found in Refs.~\cite{Lin:2021umz,Lin:2021xrc}. The total number of data points in our analysis is 1753.
\begin{table}[ht!]
	\vspace{2mm}
	\centering  
	\begin{tabular}{|c|c|c|c|}
		\hline
		\multicolumn{2}{|c|}{Data type}  &  \multicolumn{1}{c|}{range of $|t|$ [GeV$^2$]} & \multicolumn{1}{c|}{Numbers of data}   \\
		\hline
		\multirow{5}*{space-like $t<0$} & $\sigma(E,\theta)$, PRad  &  $0.000215 - 0.058$  & 71        \\
		&$\sigma(E,\theta)$, MAMI  &  $0.00384  - 0.977$     & 1422      \\
		&$\mu_p G_E^p/G_M^p$, JLab  &  $1.18 - 8.49$     & 16        \\
		&$G_E^n$, world            &  $0.14 - 1.47$     & 25        \\
		&$G_M^n$, world            &  $0.071- 10.0 $     & 23        \\
		\hline
		\multirow{4}*{time-like $t>0$} & $|G_{\rm eff}^p|$, world            &  $3.52- 20.25 $     & 153        \\
		&$|G_{\rm eff}^n|$, world            &  $3.53- 9.49 $     & 27        \\
		&$\abs{G_E/G_M}$, BaBar            &  $3.52- 9.0 $     & 6        \\
		&$d\sigma/d\Omega$, BESIII            &  $1.88^2- 1.95^2$     & 10        \\
		\hline
	\end{tabular}
	\caption{Data sets included in the combined space- and timelike fits.
		See Refs.~\cite{Lin:2021umz,Lin:2021xrc} for explicit references.}
	\label{tab:dbase}
	\vspace{-3mm}
\end{table}

Finally, let's briefly introduce the fitting strategy used in our analysis. Two different $\chi^2$ functions, $\chi^2_1$ and $\chi^2_2$, are used to measure the quality of the fits
\begin{align}
	\chi^2_1 &= \sum_i\sum_k\frac{(n_k C_i - C(t_i,\theta_i,\vec{p}\,))^2}{(\sigma_i+\nu_i)^2}~,
	\label{eq:chi1}\\
	\chi^2_2 &= \sum_{i,j}\sum_k(n_k C_i - C(t_i,\theta_i,\vec{p}\,))[V^{-1}]_{ij} \times (n_k C_j - C(t_j,\theta_j,\vec{p}\,))~,
	\label{eq:chi2}
\end{align}
where $C_i$ are the experimental data at the points $t_i,\theta_i$ and
$C(t_i,\theta_i,\vec{p}\,)$ are the theoretical value for a given FFs parametrization
for the parameter values contained in $\vec{p}$\footnote{For total cross sections and form factor data the dependence on $\theta_i$ is dropped}. Moreover, the $n_k$ are normalization
coefficients for the various data sets (labeled by the integer $k$ and only used in the fits to
the differential cross section data in the spacelike region), while $\sigma_i$ and $\nu_i$ are 
their statistical and systematical errors, respectively. The covariance matrix is $V_{ij} = \sigma_i\sigma_j\delta_{ij} + \nu_i\nu_j$. In practice,
$\chi^2_2$ is used for those experimental data where statistical and systematical errors are given separately, otherwise 
$\chi^2_1$ is adopted. As done in Ref.~\cite{Lin:2021umk,Lin:2021umz} the various constraints on the form factors are imposed in a soft way. Theoretical errors will be calculated on the one hand using the bootstrap method. On the other hand theoretical errors are estimated by varying the number of effective vector meson poles in the spectral functions. For more details on the error estimation we refer to Ref.~\cite{Lin:2021umz}.

\section{Results and Discussions}
Learning from our previous investigations~\cite{Lin:2021umk,Lin:2021umz}, the space-like data can be described quite well with those spectral functions which include all two-body continua and some additional narrow effective vector poles. Then we need to explore what is the case for the time-like data. We start with fits to the timelike data only. Note that the nucleon form factors have a non-vanishing imaginary part in the time-like region and it is found that if one wants to mimic the imaginary part of the FFs in the 
time-like region, one can e.g. allow for a non-zero width for the high
mass effective pole, that is, one needs the broad effective poles as mentioned above (see, e.g., Ref.~\cite{Belushkin:2006qa}). In the fits only to the time-like data, we vary the number of broad poles both in isoscalar and isovector spectral functions to search the best configuration for the FFs. It turns out that the structure that includes $3s+3v$ below-threshold narrow poles\footnote{Here, $3s+3v$ indicates that 3 isoscalar plus 3 isovector effective poles are included in the spectral functions. } and $3s+3v$ above-threshold broad poles serves as the best fit with $\chi^2/{\rm d.o.f}=0.638$ and can describe the observed oscillatory behavior of the effective form factors from BaBar and BESIII. 

\begin{figure}[htbp]
	\centering
	\includegraphics[width=0.47\textwidth]{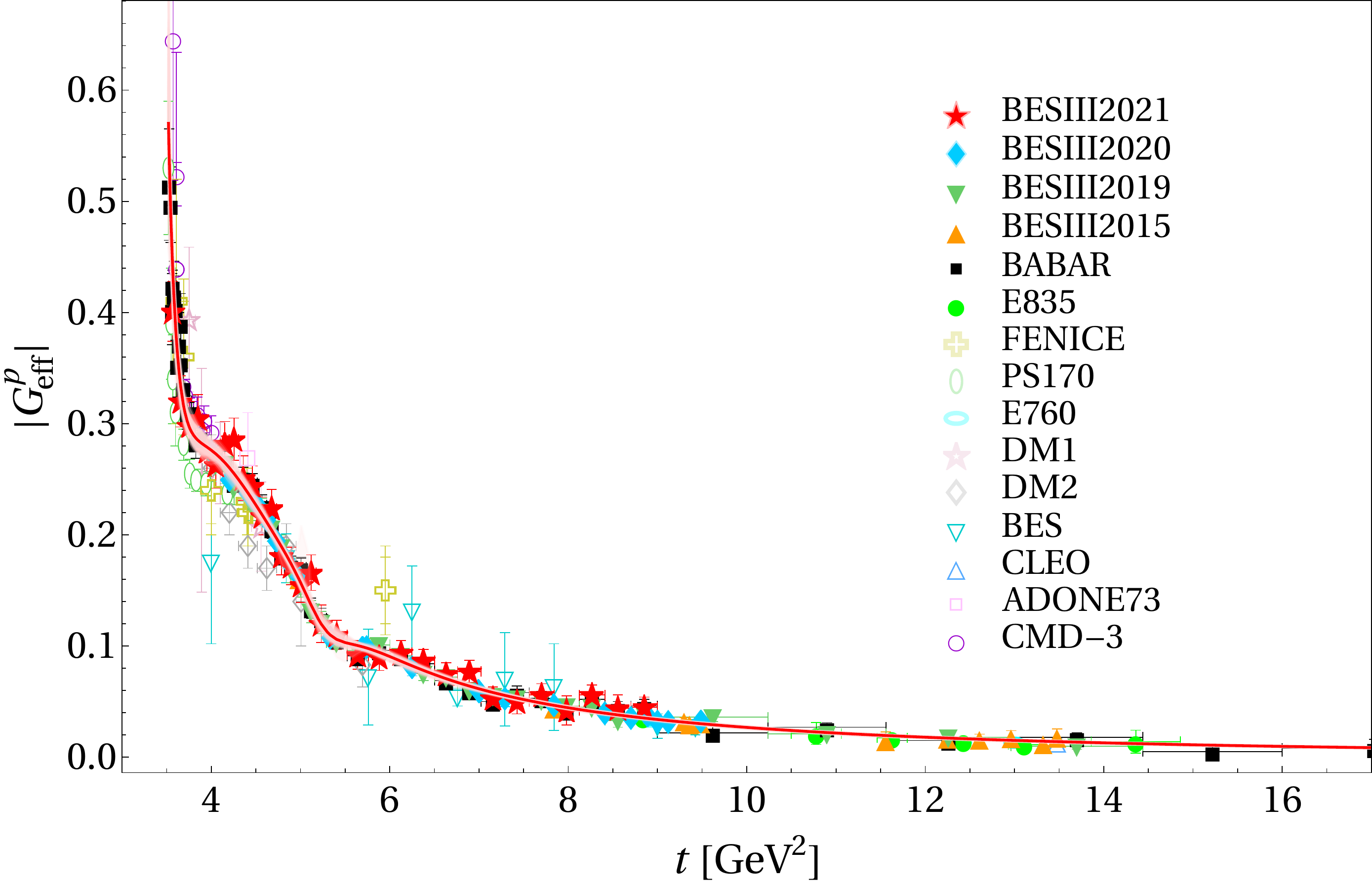}\
	\includegraphics[width=0.47\textwidth]{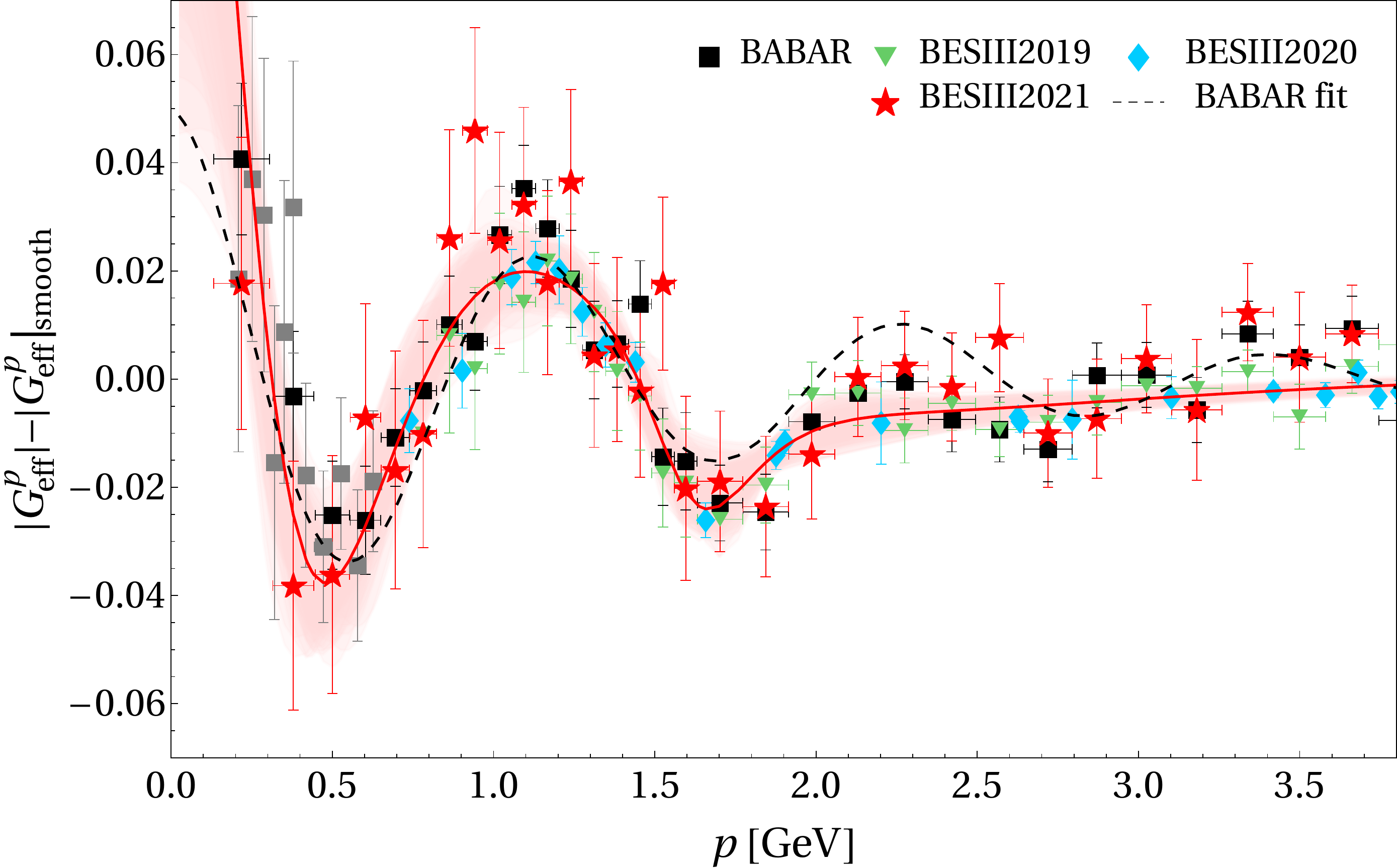}
	\caption{Complete fit to space- and timelike data
		with bootstrap error (shaded band)
		compared to data for $|G_{\rm eff}|$
		of the proton (left panel) and the oscillatory behavior in detail (right panel).
		Fitted data are depicted by closed symbols; data given
		by open symbols are shown for comparison
		only (see Ref.~\cite{Lin:2021xrc} for explicit references). $\left| G_{\rm eff}^p\right|_{\rm smooth}=\frac{7.7}{(1+t/14.8)(1-t/0.71)^2}$. The black dashed line in the right plot show the phenomenological fit to BABAR data with the formula $F_p\equiv \left| G_{\rm eff}^p\right|-\left| G_{\rm eff}^p\right|_{\rm smooth}=A \exp(-B p)\cos(C p+D)$ that proposed in Ref.~\cite{Bianconi:2015owa}. Here, $p$ is the relative momentum of the proton.}
	\label{Fig: geffp}
\end{figure}

\begin{figure}[htbp]
	\centering
	\includegraphics[width=0.47\textwidth]{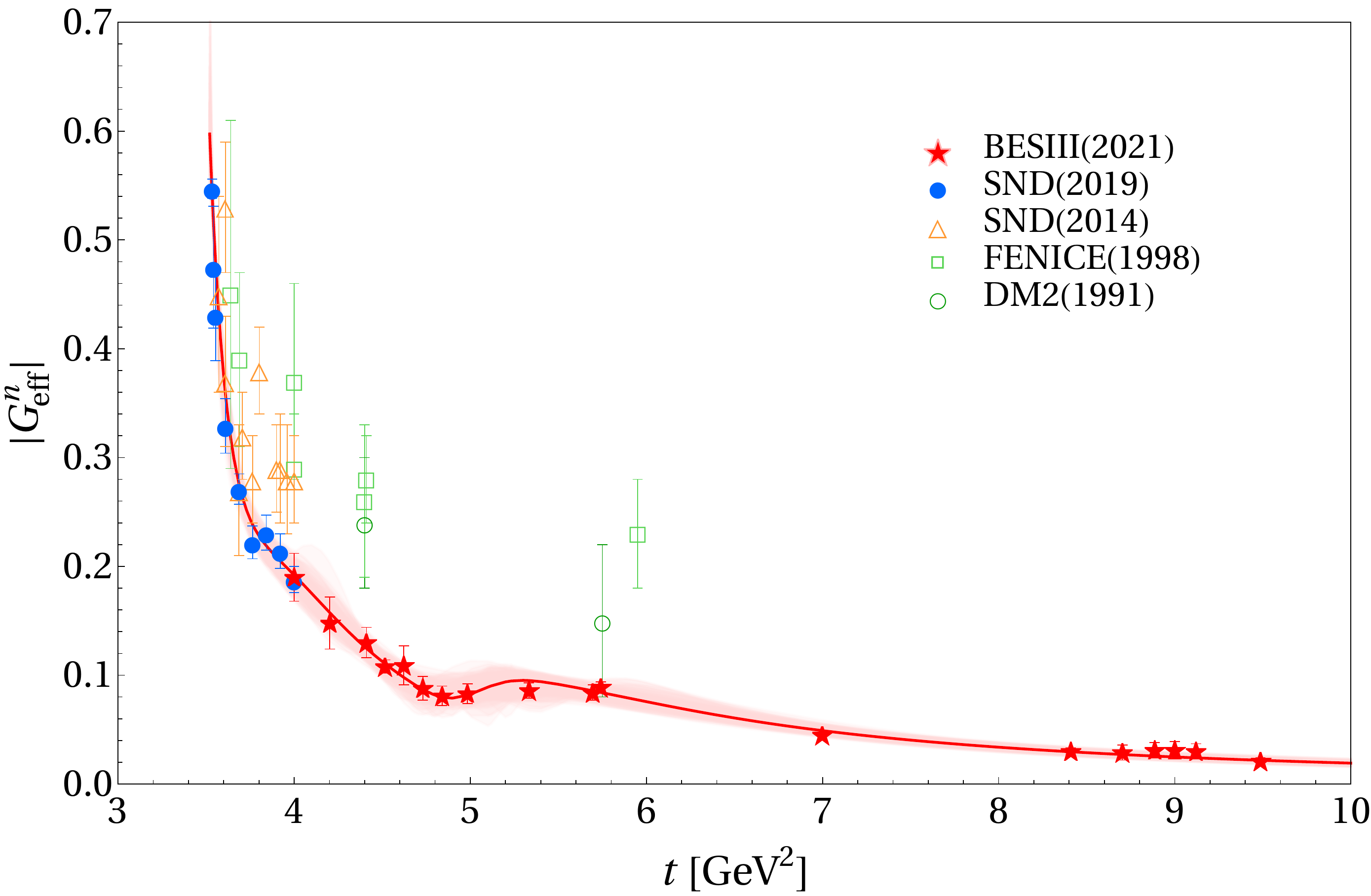}\
	\includegraphics[width=0.47\textwidth]{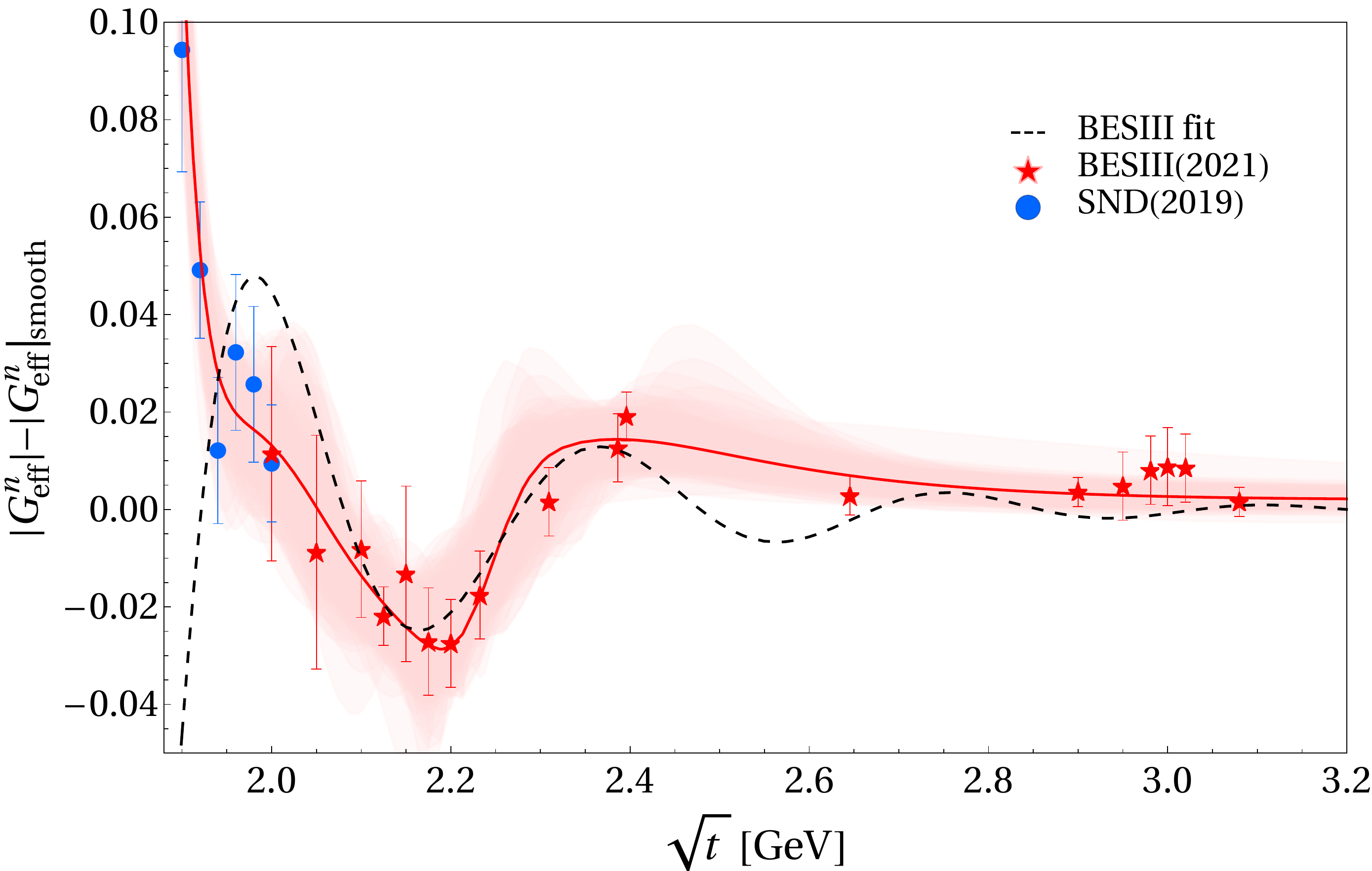}
	\caption{Complete fit to space- and timelike data
		with bootstrap error (shaded band)
		compared to data for $|G_{\rm eff}|$
		of the neutron (left panel) and the oscillatory behavior in detail (right panel).
		Fitted data are depicted by closed symbols; data given
		by open symbols are shown for comparison
		only (see Ref.~\cite{Lin:2021xrc} for explicit references). $\left| G_{\rm eff}^n\right|_{\rm smooth}=\frac{4.87}{(1+t/14.8)(1-t/0.71)^2}$~\cite{BESIII:2021dfy}. The black dashed line in the right plot show the phenomenological fit to BESIII data with the formula $F_n\equiv \left| G_{\rm eff}^n\right|-\left| G_{\rm eff}^n\right|_{\rm smooth}=A \exp(-B p)\cos(C p+D)$ that proposed in Ref.~\cite{Bianconi:2015owa}, with $p$ is the relative momentum of the neutron.}
	\label{Fig: geffn}
\end{figure}

\begin{figure}[htbp]
	\centering
	\includegraphics[width=0.47\textwidth]{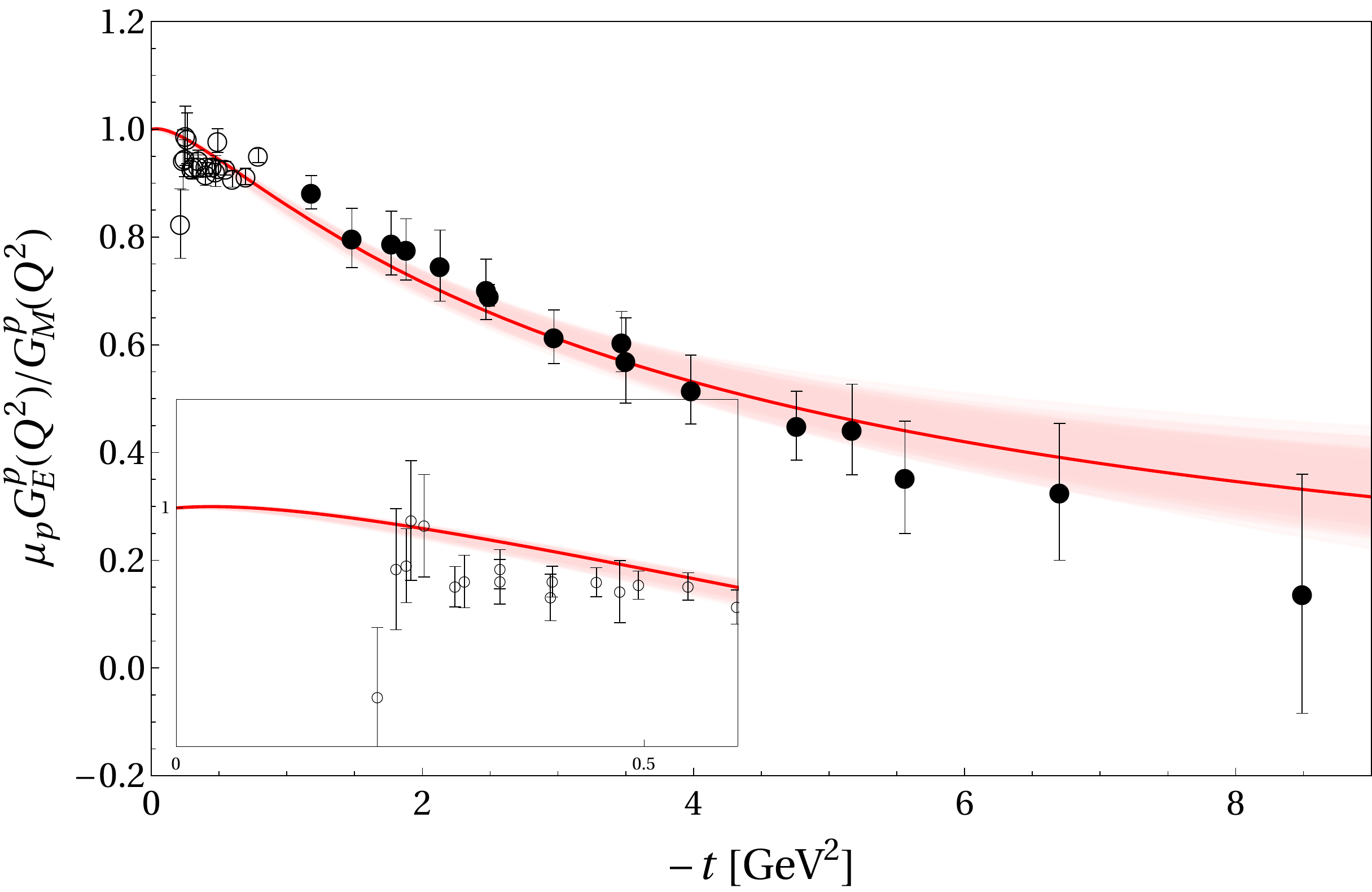}\
	\includegraphics[width=0.47\textwidth]{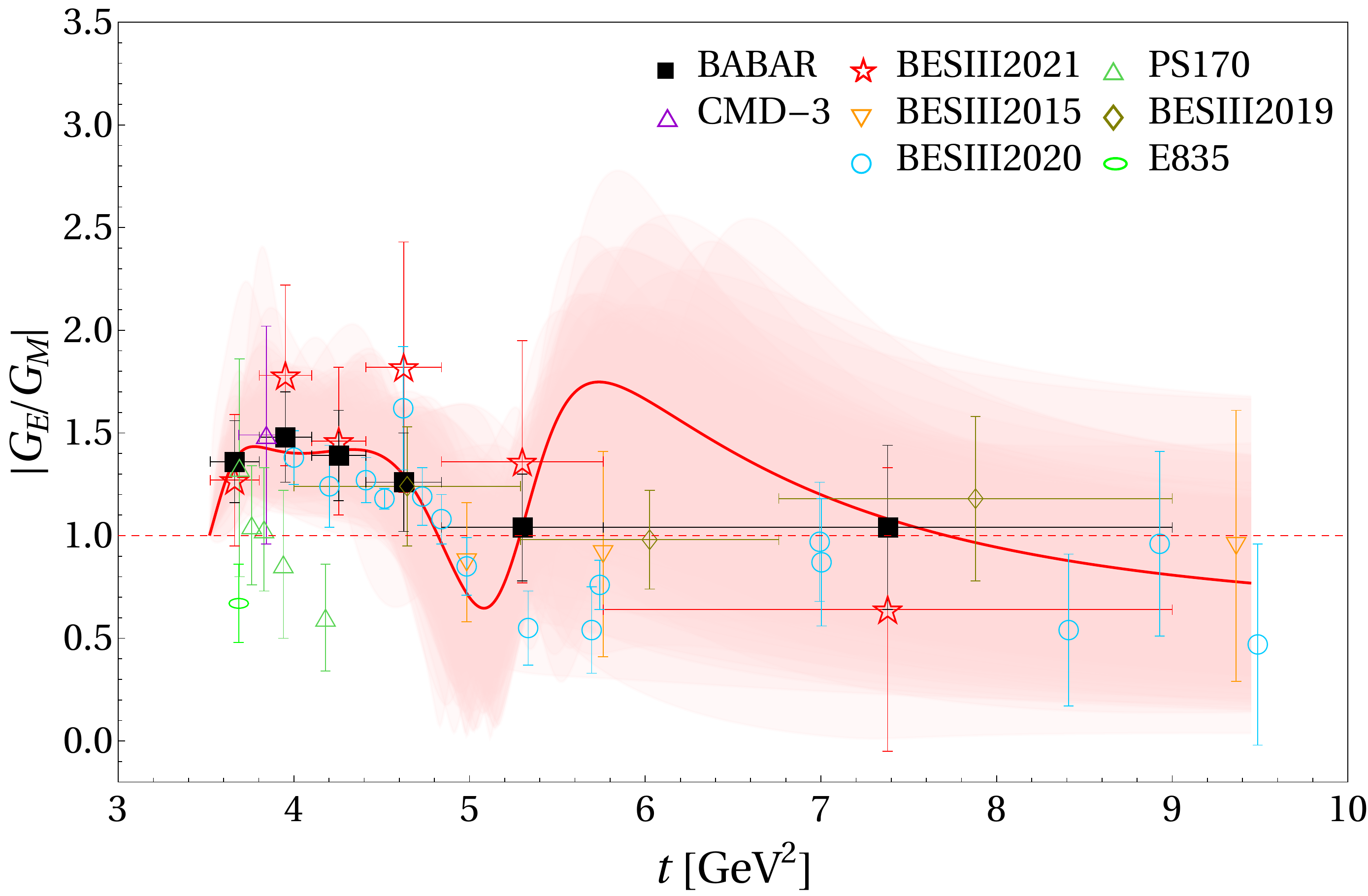}
	\caption{Complete fit to space- and timelike data
		with bootstrap error (shaded band) compared to proton
		data for $\mu_p G_E^p/G_M^p$
		at spacelike momentum transfer from JLab and for
		$\abs{G_E/G_M}$ at timelike momentum transfer (closed symbols: fitted, open symbols: not fitted). For references to the data,
		see Refs.~\cite{Lin:2021umz,Lin:2021xrc}.
		\label{Fig: geffpR}}
\end{figure}

Then we are ready to the combined fits to both space- and timelike data. To find the best fit, we only vary the number of below-threshold narrow poles while keep the number of above-threshold broad poles unchanged. The best fit is found to contain $3s+5v$ below-threshold narrow poles and $3s+3v$ above-threshold broad
poles. Remember that all two-body continua, namely the $\pi\pi$ in the isovector sector and the $\rho\pi$, $\bar{K}K$ in the isoscalar sector, are always kept. 
The quality of the fit to the spacelike data is comparable to our
previous fits of spacelike data only \cite{Lin:2021umk,Lin:2021umz}.
We obtain $\chi^2/{\rm d.o.f}=1.223$ for the full data set,
$\chi^2/{\rm d.o.f}=1.063$ for the timelike data, and 
$\chi^2/{\rm d.o.f}=1.297$ for the spacelike data.
In Fig.~\ref{Fig: geffp} and \ref{Fig: geffn}, we show our best fit compared to
the experimental data for $|G_{\rm eff}|$ of the proton and the neutron, respectively.
A similar comparison for $\mu_p G_E^p/G_M^p$ in the space-like region and $\abs{G_E/G_M}$ in the time-like region is given in Fig.~\ref{Fig: geffpR}. We would like to point out that a zero crossing of the FF ratio is disfavored by the combined analysis of space- and timelike data.

\begin{figure}[htbp]
	\centering
	\includegraphics[width=0.9\textwidth]{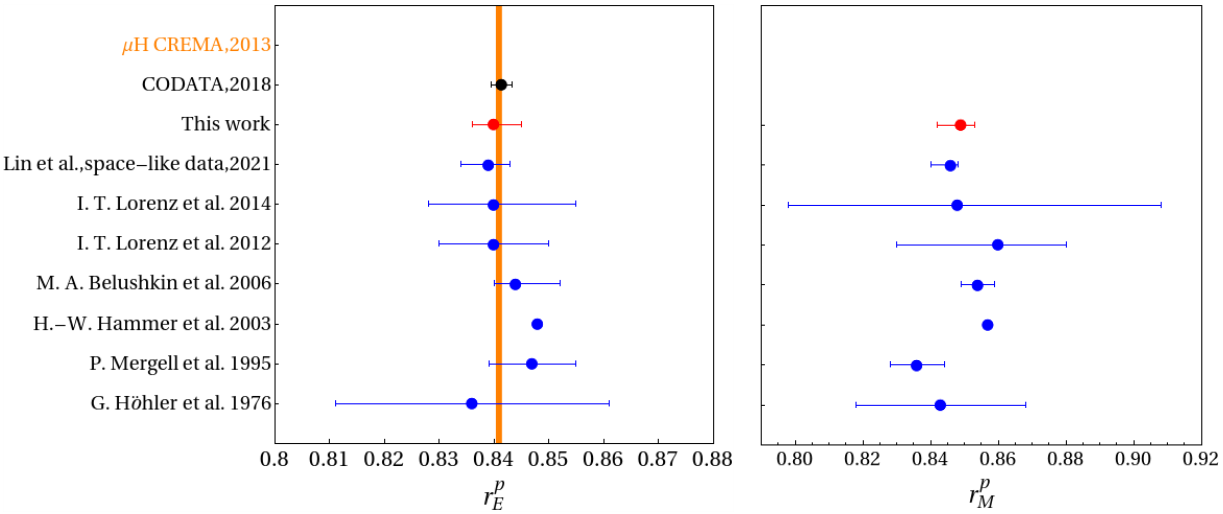}
	\caption{Comparison of the proton radii extracted in this work and other previous dispersion-theoretical extraction. Left y-axis represents the date and author of the corresponding work, see Ref.~\cite{Lin:2021umz} for the relevant papers. The orange band shows the latest radius extraction from the muonic hydrogen~\cite{Antognini:2013txn}.
		\label{fig:rpe}}
	\vspace{-3mm}
\end{figure}

\begin{figure}[htbp]
	\centering
	\includegraphics[width=0.9\textwidth]{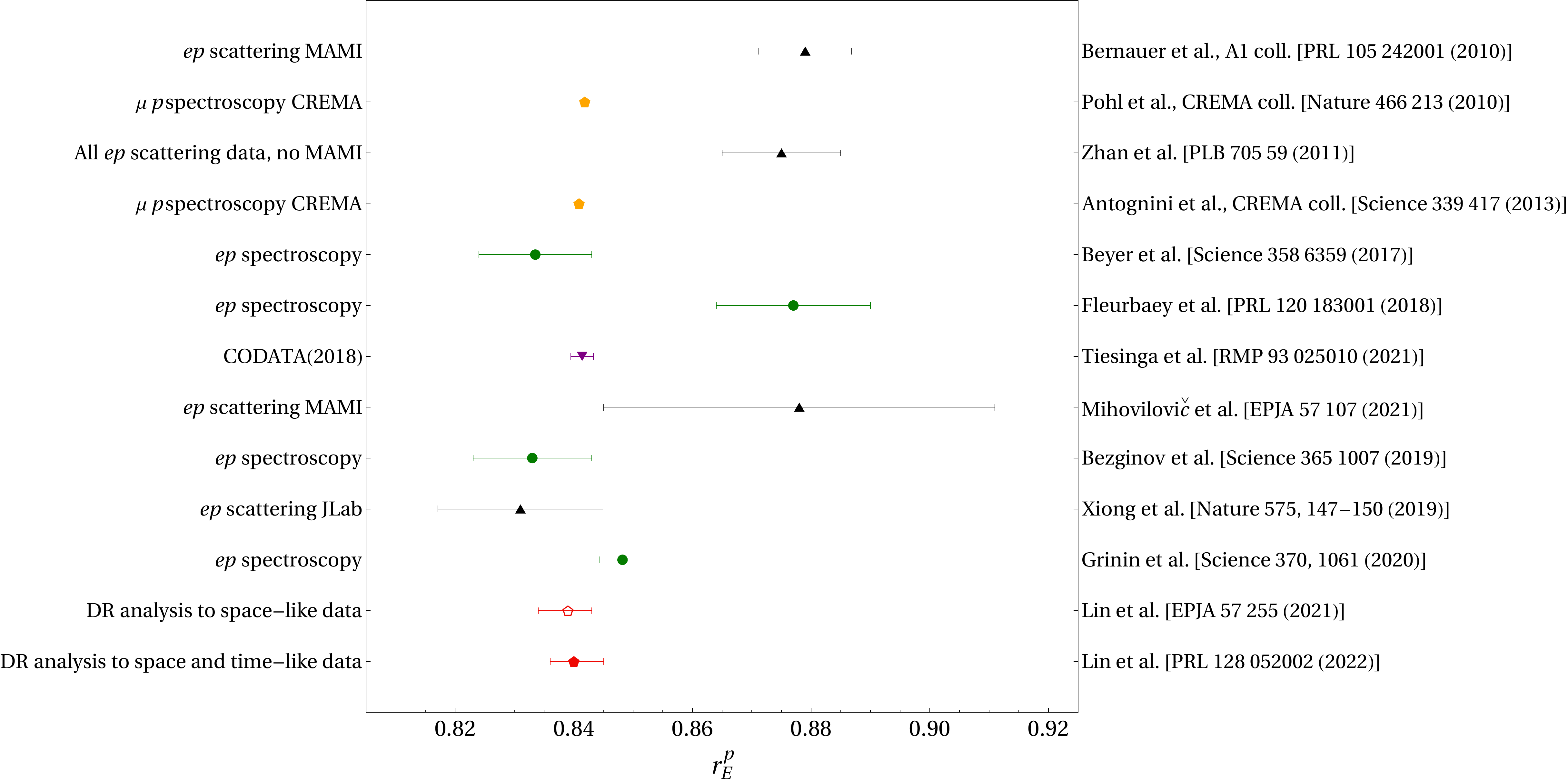}
	\caption{Comparison of the proton radii extracted in this work and other recent determinations. Left y-axis represents the process where the proton radius is extracted and right y-axis shows the corresponding reference.
		\label{fig:rpeoverview}}
	\vspace{-3mm}
\end{figure}

Now, let us move to the nucleon radii. The radii extracted from the combined fits are
\begin{eqnarray}
	r_E^p &=& 0.840^{+0.003}_{-0.002}{}^{+0.002}_{-0.002}\,{\rm fm}, \nonumber\\
	r_M^p &=& 0.849^{+0.003}_{-0.003}{}^{+0.001}_{-0.004}~{\rm fm}, \nonumber\\
	r_M^n &=& 0.864^{+0.004}_{-0.004}{}^{+0.006}_{-0.001}~{\rm fm},
	\label{eq:radii_fin}
\end{eqnarray}
where the first error is statistical (based on the bootstrap procedure) and the second one is systematic (based on the variations in the spectral functions).
These values are in good agreement with previous high-precision analyses
of spacelike data alone \cite{Lin:2021umk,Lin:2021umz} and
have comparable errors. In Fig.~\ref{fig:rpe}, we compare various dispersion-theoretical extractions from a historical perspective. Note that here we only consider those dispersion-theoretical analyses that include the two-pion continuum explicitly in their spectral functions that is found to play a crucial role in the nucleon isovector form factors, see Refs.~\cite{Lin:2021umk,Lin:2021umz} for the details. It sticks out a mile that the dispersion-theoretical analysis provides a consistent and robust proton charge radius that are in agreement with the value measured from muonic hydrogen~\cite{Antognini:2013txn}. In Fig.~\ref{fig:rpeoverview}, we list all the  recent determinations on the proton electric radius. It is shown that the agreement on the proton charge radius has already been achieved by the measurements from $ep$ scattering, $ep$ spectroscopy and the $\mu p$ spectroscopy and finally the value collected in the CODATA is updated to $0.8414(19)$\,fm~\cite{CODATAnew} which agrees quite well with our determination.

\section{Summary}

In this work, we have presented a consistent picture of the nucleons
electromagnetic structure based on all spacelike and timelike
data from electron scattering and electron-positron annihilation
(and its reversed process) for the first time. It is achieved by means of the powerful dispersion theoretical parametrization on the nucleon FFs. In particular, our best phenomenological proton form factors produce the small proton charge radius, $r_E^p=0.84\,$fm, that is consistent with earlier
dispersive analyses~\cite{Lin:2021umz} and also most recent determinations
from electron-proton scattering as well as the Lamb shift in electronic
and muonic hydrogen (as listed e.g. in Ref.~\cite{Hammer:2019uab}).

\section*{Acknowledgements}
I would like to thank Hans-Werner Hammer and Ulf-G.~Mei{\ss}ner for a most enjoyable collaboration, and specially thank Ulf-G.~Mei{\ss}ner for a careful reading. This work is supported in
part by the Deutsche Forschungsgemeinschaft (DFG, German Research
Foundation) and the NSFC through the funds provided to the Sino-German Collaborative  
Research Center TRR~110 ``Symmetries and the Emergence of Structure in QCD''
(DFG Project-ID 196253076 - TRR 110, NSFC Grant No. 12070131001), by the Chinese Aca\-de\-my of Sciences (CAS) through a President's International Fellowship Initiative (PIFI) (Grant No. 2018DM0034), by the VolkswagenStiftung (Grant No. 93562), and by the EU Horizon 2020 research and innovation programme, STRONG-2020 project under grant agreement No. 824093.

\end{document}